# Extended Prony Analysis on Power System Oscillation Under a Near-Resonance Condition

Tianwei Xia[1,2], Zhe Yu[1], Kai Sun[2], Di Shi[1], Zhiwei Wang[1]

[1]GEIRI North America, San Jose, CA, USA
[2]Department of Electrical Engineering and Computer Science, University of Tennessee, Knoxville, TN, USA
txia4@vols.utk.edu, zhe.yu@geirina.net, kaisun@utk.edu, di.shi@geirina.net, zhiwei.wang@geirina.net

*Abstract*— **Power system oscillations under a large disturbance often exhibit distorted waveforms as captured by increasingly deployed phasor measurement units. One cause is the occurrence of a near-resonance condition among several dominant modes that are influenced by nonlinear transient dynamics of generators. This paper proposes an Extended Prony Analysis method for measurement-based modal analysis. Based on the normal form theory, it compares analyses on transient and post-transient waveforms to distinguish a resonance mode caused by a near-resonance condition from natural modes so that the method can give more accurate modal properties than a traditional Prony Analysis method, especially for large disturbances. The new method is first demonstrated in detail on Kundur's two-area system and then tested on the IEEE 39-bus system to show its performance under a near-resonance condition.**

## I. Introduction

Power oscillations widely exist and often occur in interconnected power systems. By causes and mechanisms of oscillations, they can be classified into two types. The first one is the natural oscillations, which are caused by insufficient damping torques of generators especially under stressed system conditions. The other is forced oscillations driven by an external source that is usually not modeled by power engineers. There have been increasing studies both natural oscillations and forced oscillations in the past decades. For natural oscillations, several online signal processing or modal analysis methods such as Prony analysis [1] and subspace identification method [2] have been developed and applied by the power industry to estimate modal properties of natural modes, e.g. frequency, damping and phasing of oscillation. For forced oscillation, research has been focused on the detection and localization of the external oscillation source [3].

With increasing complexities and uncertainties in dynamics of modern power grids, nonlinear modal interferences will become more often to occur than ever especially under stress operating conditions or when subject to large disturbances. One example is that when a power grid has three or more natural oscillation modes in a near-resonance condition, the appearance and influence of an interference can be amplified during the transient period following a large disturbance due to the nonnegligible nonlinearity of the system. As a result, conventional signal processing or modal analysis techniques that assume linear system responses under a small distance may misrecognize a resonance mode that is not a true natural mode or fail to identify modal properties accurately. Therefore, for natural oscillations energized by a large disturbance, it is necessary to develop a new, more robust measurement-based modal analysis method.

This paper proposes an Extended Prony Analysis method, which is able to identify natural modes and estimate their modal properties more credibly than the conventional Prony analysis by identification of resonance modes using the normal form theory. The normal form theory and its variants can transform a nonlinear system into a formally linear or decoupled system in the sense of elimination of undesired nonlinear terms up to a desired order by a nonlinear coordinate transformation, which have been applied to nonlinear modal and stability analyses for a power system in an extended neighborhood of its equilibrium [4][5]. The proposed Extended Prony Analysis method compares analyses on transient and post-transient waveforms to identify a resonance mode caused by a near-resonance condition so that the method can give more accurate modal properties than a traditional Prony Analysis method especially for large disturbances. The method has two steps: in the first step, natural modes are identified from post-transient waveforms using conventional Prony analysis; then a resonance mode caused by near-resonance is detected from transient waveforms based on these natural modes. Because of considering the influence from such a resonance mode, the modal properties of true modes can be more accurately estimated. The rest of the paper is organized as follows: Section II reviews the normal form theory. Section III introduces the proposed Extended Prony Analysis method. Besides, a designed signal is applied to show the feasibility of this method. The case studies on Kundur's two-areas systems and the IEEE 39-bus system are presented in section IV. At last, the paper concludes in section V.

## II. Resonance Analysis by Normal Form Theory

Consider the differential equation model of a power system with *n* state variables:

$$\dot{\mathbf{x}} = \mathbf{f}(\mathbf{x}) \qquad (1)$$

This work was supported in part by SGCC Science and Technology Program under project "AI based oscillation detection and control" and contract No. SGJS0000DKJS1801231, in part by the ERC Program of the NSF and U.S. DOE under grant EEC-1041877 and in part by the NSF grant ECCS-1553863.

After applying Taylor series expansion to $\mathbf{f}(\mathbf{x})$ at a state equilibrium of interest, e.g. a specific steady-state operating condition, the system can be rewritten as

$$\Delta \dot{\mathbf{x}} = \mathbf{A}\Delta\mathbf{x} + \frac{1}{2}\left[\Delta\mathbf{x}^T \mathbf{H}_1 \Delta\mathbf{x} \quad \cdots \quad \Delta\mathbf{x}^T \mathbf{H}_n \Delta\mathbf{x}\right]^T + H.O.T \quad (2)$$

where the $\mathbf{A}$ is the Jacobian matrix, $\mathbf{H}_k$ ($k=1, \ldots, n$) is the Hessian matrix and "$H.O.T$" stands for all nonlinear terms higher than the 2nd order. In a small-signal study, the disturbance is considered small so as to ignore the 2nd order and higher order nonlinear terms. However, in normal form theory, the quadratic terms are kept to model the 2nd order nonlinearity of the system under a disturbance which is not ideally small.

Transform the state equation into its modal space to diagonalize $\mathbf{A}$, i.e. $\mathbf{y}=\Phi^{-1}\Delta\mathbf{x}$, where $\Phi$ is the modal matrix whose columns are eigenvectors of $\mathbf{A}$.

$$\dot{\mathbf{y}} = \Lambda\mathbf{y} + \mathbf{f}_2(\mathbf{y}) + H.O.T \quad (3)$$

whose each element is

$$\dot{y}_j = \lambda_j y_j + \sum_{k=1}^{n}\sum_{l=1}^{n} C_{kl}^j y_k y_l + H.O.T \quad (4)$$

The coefficient $\lambda_j$ is the eigenvalue of the Jacobian matrix and $C_{kl}^j$ is the coefficient of the 2nd order term. This system in the $y$-space can mathematically be transformed into an approximately linear system with nonlinear terms appearing only at the 3rd and higher orders. The purpose is to eliminate fake modes that are caused by the 2nd order terms in order for modal analysis on natural modes to be more accurate. This can be done by a nonlinear transformation $\mathbf{y}=\mathbf{h}(\mathbf{z})$:

$$y_j = z_j + \sum_{k=1}^{n}\sum_{l=1}^{n} h2_{kl}^j z_k z_l \quad (5)$$

$$h2_{kl}^j = \frac{C_{kl}^j}{\lambda_k + \lambda_l - \lambda_j} \quad (6)$$

Then, the final analytical solutions in $x$, $y$ and $z$ spaces are

$$z_j(t) = z_{j0} e^{\lambda_j t}$$

$$y_j(t) = z_{j0} e^{\lambda_j t} + \sum_{k=1}^{n}\sum_{l=1}^{n} h2_{kl}^j z_{k0} z_{l0} e^{(\lambda_k + \lambda_l)t} \quad (7)$$

$$x_i(t) = \sum_{j=1}^{n} u_{ij} z_{j0} e^{\lambda_j t} + \sum_{j=1}^{n} u_{ij}[\sum_{k=1}^{n}\sum_{l=1}^{n} h2_{kl}^j z_{k0} z_{l0} e^{(\lambda_k + \lambda_l)t}]$$

where $z_{j0}$, $z_{k0}$ and $z_{l0}$ denote the initial values in $z$-space. From the 3rd equations, we can clearly see that new modes corresponding to the sums of two eigenvalues appear, which are called resonance modes in the rest of the paper. These modes exist with a nonlinear system but are manifested only when a near resonance condition occurs and the disturbance drives the system state away from the equilibrium (note that the amplitude of a resonance mode depends on the product of the state deviations in $z$-space).

If the system is under a near-resonance (2nd order resonance) condition, which means that

$$\lambda_k + \lambda_l \approx \lambda_j \quad j = 1, 2, \ldots, n \quad (8)$$

then the coefficient $h2_{kl}^j$ of second order terms will be extremely large.

As a conclusion, once a resonance or near-resonance condition occurs, the original nonlinear power system modeled in $x$-space can hardly be transformed into a linear system by any nonlinear transformation due to the very large coefficients of transformation $\mathbf{y}=\mathbf{h}(\mathbf{z})$. In other words, the resonance or near-resonance condition amplifies the nonlinear characteristics of the power system near the equilibrium. Thus, the power system in such a condition should not be studied as a linear system.

## III. EXTENDED PRONY ANALYSIS

### A. Idea and steps

The idea of Prony analysis is to fit a time window of signal, e.g. a post-disturbance system response, into a set of damped sinusoidal signals [1]:

$$y(t) = \sum_{i=1}^{p} A_i e^{\alpha_i t} \cos(2\pi f_i t + \theta_i) \quad (9)$$

Or in other words, to fit an input signal $y(t)$ sampled at intervals of $T$, a set of exponential functions are determined:

$$y(kT) = \sum_{i=1}^{p} B_i e^{\lambda_i kt} = \sum_{i=1}^{p} B_i \mu_i^k \quad (10)$$

where

$$\mu_i = e^{\lambda_i t} \quad \lambda_i = \alpha_i \pm j\omega_i \quad B_i = \frac{A_i}{2} e^{j\theta_i} \quad (11)$$

For the $i$-th oscillation mode, $A_i$ represents the amplitude. $B_i$ denotes its amplitude and is called the contribution factor in [6]. Comparing the contribution factors in equations (7) and (10), the conventional Prony analysis may provide a wrong result when the second order terms of (7) are large. It will regard the resonance modes as some natural modes at higher oscillation frequencies and give inaccurate contribution factors as well as mode shapes on true natural modes.

The proposed Extended Prony Analysis assumes: (1) a properly designed power system does not have ideally resonating natural modes, i.e. "≈" in (8) instead of exact equality; (2) floating modal properties may cause a near-resonance condition in a limited transient period that does not last long. The second assumption is based on the studies in [7][8] on nonlinearity in power system oscillation, which shows that the frequency of an electromechanical mode decays with the increase of oscillation amplitude according to an F-A (frequency-amplitude) curve. Also, from our studies, we found that in most cases, a near-resonance condition may appear in the first several seconds of a transient period following a large disturbance, and then disappear.

Thus, the proposed Extended Prony Analysis method conducts analyses on waveforms from both transient (first several seconds) and post-transient periods after a disturbance. It has the following steps:

1. Divide a time window of waveform data into two segments. The first segment contains data right after the disturbance until the oscillation amplitude is sufficiently small, and the rest of the data are assigned to the second segment.

2. Apply conventional Prony analysis to the second segment to identify credible natural modes without resonance and estimate their frequencies as the reference.

3. Based on identified natural modes, possible resonance modes are constructed according to (7).

4. Fit the waveforms in the first segment into constructed resonance modes. Then the accurate contributions factor and mode shapes of the natural modes can be computed.

### B. Illustration on designed signals

Consider a signal $x(t)$ containing one non-oscillatory mode ($\lambda_1$) and three oscillatory modes ($\lambda_2$ to $\lambda_7$) given in TABLE I:

$$x(t) = \sum_{i=1}^{7} B_i e^{\lambda_i t} \quad (12)$$

The last mode 4 is assumed to have decaying frequency with the increase of amplitude according to the F-A curve in Fig. 1 for a typical electromechanical mode.

TABLE I. THE PARAMETERS OF THE DESIGNED SIGNALS

| $i$ | 1 (Mode 1) | 2, 3 (Mode 2) | 4, 5 (Mode 3) | 6, 7 (Mode 4) |
|---|---|---|---|---|
| $\lambda$ | -0.3199 | -0.1433±j3.3931 | -0.1869±j6.8812 | -0.3300±j$\omega$ |
| $B$ | 0.500 | 0.500 | 0.500 | 0.500 |
| $f$ (Hz) | - | 0.540 | 1.095 | 1.430 to 1.670 |

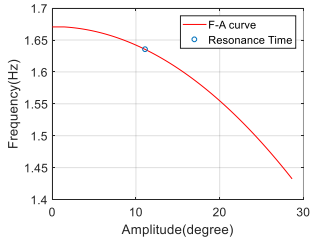

Fig. 1. The F-A curve of mode 4

When $t$ is around 3.16 s, a near-resonance condition happens between modes 1 and 2 to create a resonance mode at a frequency close to that of mode 4:

$$\lambda_6 \approx \lambda_2 + \lambda_4$$
$$f_6 \approx f_2 + f_4 = 1.635 \quad (13)$$

TABLE II compares the conventional Prony analysis with the proposed Extended Prony Analysis on data in Fig. 2 from 2 seconds to 25 seconds including a short time with obvious nonlinearity of mode 4. Considering the inability of Prony analysis in distinguishing a resonance mode from natural modes, it is respectively applied to 4 natural modes (7 eigenvalues) and to 5 modes (9 eigenvalues) including the resonance mode, which are called Strategy 1 and Strategy 2, respectively in the table. In Fig. 2, the purple curve shows the degree of resonance computed by (7). The Extended Prony Analysis (Strategy 3) divides the input data into two segments at $t$=10 seconds. The second segment is used for conventional Prony analysis, and the first segment is fitted into the constructed modes by (7).

TABLE II. THREE STRATEGIES FOR DESIGNED SIGNALS STUDY

| # | Strategy | Time window (s) | | Mode number |
|---|---|---|---|---|
| | | Start | End | |
| 1 | Prony | 2 | 25 | 7 |
| 2 | Prony | 2 | 25 | 9 |
| 3 | Extended Prony | 2 | 25 | 9 |

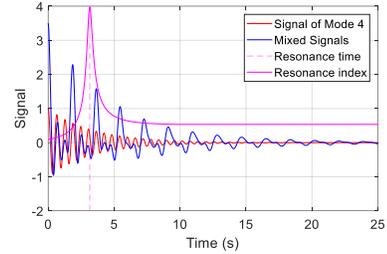

Fig. 2. The trajectories of the designed signals

The performance of the three strategies is compared in TABLE III. We can notice that the Prony analysis fails to detect the resonance mode in Strategies 1 and 2. However, the Extended Prony Analysis distinguishes the resonance mode from the natural modes. Besides, it provides more accurate contribution factors.

TABLE III. THE PERFORMANCE OF THREE STRATEGIES

| # | Strategy 1 | | Strategy 2 | | Strategy 3 | |
|---|---|---|---|---|---|---|
| | $f$ (Hz) | $B$ | $f$ (Hz) | $B$ | $f$ (Hz) | $B$ |
| 1 | 0 | 0.5855 | 0 | 0.5855 | 0 | 0.5076 |
| 2,3 | 0.868 | 0.4849 | 0.868 | 0.4849 | 0.540 | 0.5021 |
| 4,5 | 1.656 | 0.3289 | 1.660 | 0.3289 | 1.095 | 0.5032 |
| 6,7 | 40.548 | 0.0001 | 40.548 | 0.0001 | 1.671 | 0.7714 |
| 8,9 | - | | 60.241 | 0.0001 | 1.635 | 0.3649 |

## IV. CASE STUDIES ON POWER SYSTEMS

### A. Kundur's two area system

Kundur's two-area system is used for the first case study, whose topology is shown as

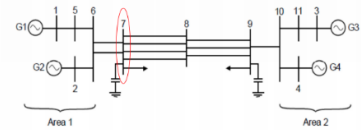

Fig. 3. Kundur's two-area system [9]

A 3-phase fault occurs on bus 7, and it is cleared after 5 cycles with no line trip. The trajectories of the four generators' rotor speeds over a time window of $t$=2-25 s are used for modal analysis using three strategies shown in TABLE IV targeting at 8 possible modes. Strategies 1 and 2 respectively apply Prony analysis to the whole time window and a post-transient portion of the window ($t$=15-25 s) while Strategy 3 applies the Extended Prony Analysis to the whole window.

TABLE IV. THREE STRATEGIES FOR TWO-AREA SYSTEM STUDY

| # | Strategy | Time window (s) Start | Time window (s) End | Mode number |
|---|---|---|---|---|
| 1 | Prony | 2 | 25 | 8 |
| 2 | Prony | 15 | 25 | 8 |
| 3 | Extended Prony | 2 | 25 | 8 |

TABLE V. shows the mode shapes of different modes after normalization. The angle and length of each phasor respectively give the phasing and amplitude of the mode. The mode shapes from the linearized system model are used as the reference. However, note that the model-based modal properties are only accurate for small-signal stability analysis.

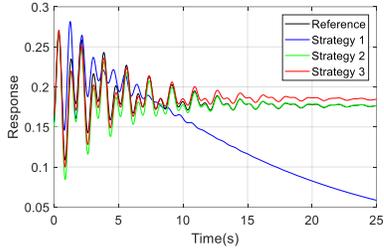

Fig. 4. The trajectories of three strategies based on the data from generator 1

The three different strategies provide different mode shapes. Since the mode information is already known from the system model, the trajectories on the modes identified by each strategy can be reconstructed in the time domain. Fig. 4 shows the results of those strategies based on the data from generator 1. The black curve is the numerical simulation by the 4th order R-K method, which is used as an accurate reference. The blue curve (Strategy 1) and the green curve (Strategy 2) are the conventional Prony analysis with different time windows. It is obvious that the green curve is the closest to the reference after 10 seconds, while the Extended Prony Analysis as shown by the red curve (Strategy 3) is the overall best. The blue curve is far from the black curve due to nonlinearity with the system response, so Prony analysis is not accurate for data after such a large disturbance. Analyses on the other three generators give similar observations and are not shown here.

Strategy 1 provides very inaccurate mode shapes because it treats all modes as natural modes. For Strategy 2, since it only applies Prony analysis to post-transient data, its result is very close to the model-based small-signal stability analysis result, especially on the dominate 0.5 Hz mode and 1.1 Hz mode. The result of the Extended Prony Analysis with Strategy 3 is slightly different on these two modes. However, on the 1.6 Hz resonance mode, the contribution from generator 2 is much larger than the model-based result as well as the result from Strategy 2.

On the 1.6 Hz mode, it can be noticed that directions of phasors in the mode shapes from Strategy 3 are not distributed in a balanced manner. It seems a phasor is missing in the 2nd quadrant for a balance. If the resonance is also considered on this mode, the new mode shape result is shown in Fig. 5 a). The purple arrow indicates the contribution of generator 4 with the resonance mode. The mode shapes on this resonance mode about the four generators are shown in Fig. 5 b). The combined contribution of four generators on the resonance mode actually behaves like the "missing" phasor in the 1.6 Hz mode shape on generator 2. The conventional Prony analysis may draw a misleading conclusion that generator 4 far more important than other generators. However, by using the Extended Prony Analysis, influence and contribution from generator 2 in the transient period can be estimated.

TABLE V. THE MODE SHAPES

| $f$ | Model-based result | Strategy 1 (Prony Analysis on 2-25 s) | Strategy 2 (Prony analysis on 15-25 s) | Strategy 3 (Extended Prony Analysis on 2-25 s) |
|---|---|---|---|---|
| 0.5Hz | | | | |
| 1.1Hz | | | | |
| 1.6Hz | | | | |

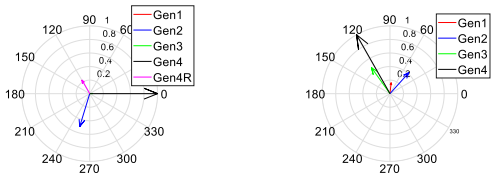

a) Mode shapes of 1.6 Hz mode   b) Mode shapes of the resonance mode

Fig. 5.  The mode shapes considering the resonance mode

## B. IEEE 39-bus system

The proposed method is also tested on the IEEE 39-bus system shown in Fig. 6 with a temporary three-phase fault on bus. Fig. 7 shows the frequency response of the generator on bus 35 in the time domain respectively after a small disturbance and a large disturbance.

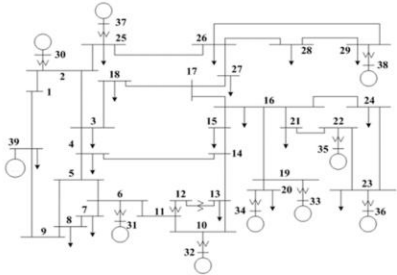

Fig. 6.  IEEE 39-bus system

Compare two strategies both considering 10 modes: 1) the conventional Prony analysis uses the data from 15 seconds to 30 seconds to avoid the transient period, and 2) the Extended Prony Analysis uses the data from $t$=3 to 30 seconds divided at $t$=15 seconds. Fig. 7 also plots the reconstructed curves from two strategies.

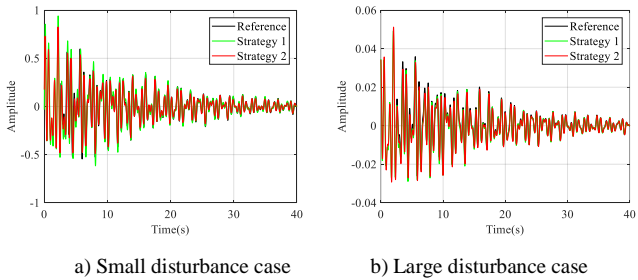

a) Small disturbance case   b) Large disturbance case

Fig. 7.  The response and reconstructed response of the generator on bus 35

In order to measure the accuracy of those two methods, we introduce the error index defined as the percentage error between the integrated reference result and reconstructed signal by the tested method:

$$e_{tr} = \frac{\int_0^T |x_{signal}(t) - x_{ref}(t)| dt}{\int_0^T |x_{ref}(t)| dt} \times 100\% \quad (14)$$

Firstly, consider a small disturbance, which only last 2 cycles. TABLE VI. shows the errors for different time intervals. The conventional Prony analysis is more accurate since it avoids using data in the transient period and also resonance is not serious in this case. For the Extended Prony Analysis, it faces more fitting challenges due to considering additional resonance modes, so it has slightly bigger but still acceptable errors.

For a large disturbance, the fault duration is prolonged to 0.26 seconds, which is closed to the critical clear time. Using conventional Prony analysis over different time windows, we could detect a 3.383 Hz mode, which is only strong in the first 6 seconds. In fact, it is found by the Extended Prony Analysis to be a resonance mode having its frequency close to the sum of frequencies of the first two dominant modes.

TABLE VI. also shows the errors at different time intervals. The conventional Prony analysis is more accurate only after $t$=20 seconds, while the Extended Prony Analysis is more reliable over the entire time window.

TABLE VI.   THE ERRORS INDEX OF DIFFERENT STRATEGIES AND CASES

| Time window (s) | | 2.15-35 s | 2.15-20 s | 20-35 s |
|---|---|---|---|---|
| Small disturbance | Strategy 1 | $1.966\times10^{-5}$ % | $3.033\times10^{-5}$ % | $6.951\times10^{-6}$ % |
| | Strategy 2 | $3.688\times10^{-5}$ % | $5.338\times10^{-5}$ % | $1.725\times10^{-5}$ % |
| Large disturbance | Strategy 1 | 0.075% | 0.126% | 0.014% |
| | Strategy 2 | 0.062% | 0.086% | 0.034% |

## V. CONCLUSION

This paper has proposed an Extended Prony Analysis method for a more reliable measurement-based modal analysis of a power system under a large disturbance. It is based on the normal form theory and considers modal interferences in near-resonance conditions. The conventional Prony analysis and the proposed Extended Prony Analysis are compared in detail on test power systems to demonstrate the good performance of the new method. The improvement of the new method in accuracy under small disturbances can be a focus of our future work.

REFERENCES

[1] J. F. Hauer, C. J. Demeure, L. L. Scharf, "Initial results in Prony analysis of power system response signals," IEEE Transactions on Power Systems, vol. 5, no. 1, 80-89, 1990.

[2] Z. Ning, J. W. Pierre, and J. F. Hauer, "Initial results in power system identification from injected probing signals using a subspace method," IEEE Transactions on Power Systems, vol. 21, no. 3, 1296-1302, 2006.

[3] B. Wang, K. Sun, "Location Methods of Oscillation Sources in Power Systems: A Survey," Journal of Modern Power Systems and Clear Energy, vol. 5, No. 2, pp. 151-159, 2017

[4] L. Shu, A. R. Messina, and V. Vittal, "Assessing placement of controllers and nonlinear behavior using normal form analysis," IEEE Transactions on Power Systems, vol. 20, no. 3, 1486-1495, 2005.

[5] B. Wang, K. Sun, W. Kang, "Nonlinear modal decoupling of multi-oscillator systems with applications to power systems," *IEEE Access*, vol.6, pp.9201-9217, 2018.

[6] M. Netto, Y. Susuki, L. Mili, "Data-Driven Participation Factors for Nonlinear Systems Based on Koopman Mode Decomposition," IEEE Control Systems Letters, vol. 3, no. 1, 198-203, 2019.

[7] B. Wang, K. Sun, "Formulation and characterization of power system electromechanical oscillations," *IEEE Trans. Power Syst.*, vol.31, no.6, pp.5082-5093, 2016.

[8] B. Wang, X. Su, K. Sun, "Properties of the Frequency–Amplitude Curve," IEEE Transactions on Power Systems, vol. 32, no. 1, 826-827, 2017.

[9] P. Kundur, Power System Stability and Control, 1993